\documentclass[fleqn,usenatbib]{mnras}

\usepackage{newtxtext,newtxmath}

\usepackage[T1]{fontenc}

\DeclareRobustCommand{\VAN}[3]{#2}
\let\VANthebibliography\thebibliography
\def\thebibliography{\DeclareRobustCommand{\VAN}[3]{##3}\VANthebibliography}

\usepackage{graphicx}	
\usepackage{amsmath}	
\usepackage{ulem}
\usepackage{comment}
\usepackage{xcolor}
\usepackage{orcidlink}

\title[Stellar winds with strongly localized heating]{Polytropic stellar wind models with strongly localized heating}

\author[L. Westrich et al.]{
L. Westrich $^{1,2,}$\orcidlink{0009-0000-2739-5418}\thanks{E-mail: Lukas.Westrich@ruhr-uni-bochum.de},
B. Shergelashvili $^{1,2,3,4,}$\orcidlink{0000-0001-7179-466X},
H. Fichtner $^{1,}$\orcidlink{0000-0002-9151-5127},
and V. N. Melnik$^{5}$
\\
$^{1}$Theoretical Physics IV, Ruhr-Universität Bochum, Universitätsstrasse 150, 44780 Bochum, Germany\\
$^{2}$Centre for Computational Helio Studies, Faculty of Natural Sciences and Medicine, Ilia State University, Cholokashvili Ave. 3/5, 0162 Tbilisi, Georgia\\
$^{3}$Evgeni Kharadze Georgian National Astrophysical Observatory, M. Kostava street 47/57, 0179 Tbilisi, Georgia\\
$^{4}$Institut f\"ur Weltraumforschung, \"Osterreichische Akademie der Wissenschaften, Schmiedlstrasse 6, 8042 Graz, Austria\\ 
$^{5}$Institute of Radio Astronomy, National Academy of Sciences of Ukraine, Kharkov 61002, Ukraine
}

\date{Accepted XXX. Received YYY; in original form ZZZ}

\pubyear{\the\year{}}

\begin{document}
\label{firstpage}
\pagerange{\pageref{firstpage}--\pageref{lastpage}}
\maketitle

\begin{abstract}
Polytropic models of stellar winds remain to be useful tools because they allow
for a simple description of the energy balance of the expanding plasma without 
explicitly specifying potentially complex energy transport processes like, 
e.g., heat conduction or extended wave heating. Among recent applications to 
stellar winds and to the solar wind was a study of the consequences of strongly
localized heating in the latter, possibly due to acoustic waves. Such 
'nonuniform' heating can result from a
time- and space-localized damping of wave modes and allows, as an extreme case, 
an adiabatic expansion of particular wind streams outside the heating region.
The present study generalizes the modeling from the first analytical as well
as numerical studies, that were limited to this extreme case, towards a more
realistic non-adiabatic behaviour. The additional energy due to heating is demonstrated to be in a plausible range in view of typical flare energies and low compared to the gravitational energy of the plasma in this region. The corresponding solutions may be of 
interest for stellar winds, in general, and w.r.t.\ recent observations made 
with the Parker Solar Probe, which revealed strongly varying wind streams and
the presence of acoustic waves near the Sun, for the solar wind, in particular. Potential observational evidence for the solar wind is discussed.
\end{abstract}




\section{Introduction} \label{sec:intro}
A useful tool for the modeling of stellar winds is the polytropic equation of state \citep[see, e.g.,][]{Parker-1960,Kopp-Holzer-1976,Lamers-Cassinelli-1999}, because it allows for a simple closure of the system of fluid equations \citep[e.g.,][]{Vidotto-etal-2015,Cohen-2017}, and, simultaneously, for an implicit treatment of an extended 'uniform' heating and cooling of the expanding plasma \citep[e.g.][]{Meskini-etal-2024}. Particularly, the formulation in the latter paper shows that heating and cooling may be modeled with an effective polytropic index $\alpha < 5/3$ and $\alpha > 5/3$, respectively. Such effective polytropic index can be derived from measurements, see for recent examples \citet{Nicolaou-etal-2020}, \citet{Pang-etal-2020}, \citet{Shaikh-etal-2023}, or \citet{Katsavrias-etal-2025} that all analyzed solar wind data. 

Employing the polytropic approach in the adiabatic limit, \citet{Shergelashvili-etal-2020} have demonstrated, recently, that within the framework of Parker's classical treatment \citep{Parker-1960} a new class of solutions can be described when allowing for a strongly localized heating. The latter, which may possibly be due to acoustic waves as discussed in \citet{Shergelashvili-etal-2020}, can transform rather slow subsonic streams into supersonic ones. These authors showed that in the case of an extremely narrow heating region at the critical (sonic) point the resulting solar wind solution can be constructed semi-analytically as a previously unnoticed, but physically meaningful 'discontinuous' combination of sub- and supersonic branches of the classical Parker model. A first application of the new solutions, which exhibit strong density depletions, has been made \citep{Shergelashvili-etal-2020} so-called unusual type-III radio bursts \citep{Melnik-etal-2014}, whose cut-offs appear to be consistent with the related strong density gradients. Such drastically variable  solar wind profiles can be important in the general framework of various values of the polytropic index $\alpha$ for the understanding of other Type-III radio bursts with sign reversal of the spectral drift rate \citep{Melnik2015}. 

Such solar wind solutions may also be of interest w.r.t.\ recent results from the Parker Solar Probe \citep[PSP,][]{Fox-etal-2016}. One of the major findings of the PSP is a high variability of the solar wind close to the Sun that is especially observed for the slow solar wind \citep{Magdalenic-etal-2023, Raouafi-etal-2023}. As \citet{Schwanitz-etal-2021} point out, the origin of the complex variability is not yet understood but may result from varying strengths of sources feeding the wind. This variability includes extreme cases like a near-subsonic outflow exhibiting a massive density depletion and a very low wind speed of around 150~km/s out to beyond 15 solar radii \citep{Ervin-etal-2024}. Another finding is the existence of ion-acoustic waves near the Sun \citep{Mozer-etal-2021, Afify-etal-2024} that have been shown to be able to heat and accelerate the solar wind \citep{Kellogg-etal-2024}. 

Because these observations are potentially in line with the model discussed in \citet{Shergelashvili-etal-2020}, \citet{westrich2024} relaxed the assumption of an implicitly assumed delta-function heating to the physically more realistic case of a somewhat extended, but still localized, heating region and considered 'quasi-discontinuous' solar wind solutions. By using both steady-state and time-dependent modeling, it was demonstrated that indeed in the limiting case of a delta-function heating the results of \citet{Shergelashvili-etal-2020} are reproduced. It may be interesting to note that, recently, also invoking ad-hoc heating sources but additionally heat conduction and its dissipation, \citet{Song-etal-2025} have picked up the idea of 'discontinuous' solutions for the solar wind expansion, demonstrating that their existence is not a consequence of the polytropic assumption. It should also be mentioned that the property of the solar wind pattern is stable against isolated distortions in the absence of additional heat or energy sources, which has been analytically shown by \citet{Shivamoggi2023}, while at the same time the generically unsteady wind evolutions are also possible \citep{Shivamoggi2025}.

Given that in view of the above-mentioned PSP findings, such approach appears promising to model some of the solar wind streams that in their totality constitute the observed variability. This, however, requires to improve the modeling such that not only principal but rather refined solutions can be studied that allow for a more realistic interpretation of actual observations. Therefore, we extend the modeling with the present paper from a (quasi-)adiabatic to a non-adiabatic expansion outside the heating region, this way allowing for the presence of a simultaneous 'uniform' heating of the wind, e.g., due to Alfv\'en waves \citep[e.g.,][]{Tu-Marsch-1995, Shergelashvili-Fichtner-2012} on top of which a localized heating, possibly due to acoustic waves, takes place. Interestingly, despite a few studies on time-localized heating
\citep[e.g.,][]{Lie-Svendsen-etal-2002} and spatially localized heating \citep[e.g.,][]{Osman-etal-2012}, the effect of local and, thus, 'nonuniform' heating (as termed by the latter authors) appears to be not extensively studied, so far. 

The paper is organized as follows. In the following section~2 the effect of local heating of a non-adiabatic polytropic wind is studied analytically implicitly assuming a delta-function like heating. After a discussion of the energy budget required for the local heating and of the constraints on the solutions from momentum conservation in Section~3, in Section~4 numerical solutions of the wind equations allowing for an extended heating region are presented. Section~5 contains a discussion of the heuristic basis of using generalized polytropic indices, in Section~6 potential observational evidence is discussed and, finally, in Section~7 all results and conclusions are summarized. 
\section{Discontinuous-type 1D solutions with \(\alpha \neq 5/3\)} \label{sec:SWarbitraryadiab}
\subsection{Analytical model approach}
In this subsection, we present a general derivation of a stationary solar wind model with discontinuous profiles for arbitrary values of the (generalized) polytropic index (or exponent) $\alpha$ (see section \ref{sec:heu}). Keeping the same notation, we start from the basic set of differential equations which have been derived in \citet{Shergelashvili-etal-2020}:
\begin{equation}
\left(5-3\alpha\right)\frac{1}{C_{s}^{2}}\frac{dC_{s}^{2}}{dr}+\left(\alpha-1\right)\left(4\frac{d\ln\eta}{dr}+\frac{d\ln\xi}{dr}\right)=0,\label{eq: soundgeneral}
\end{equation}
\begin{equation}
\frac{d\xi}{dr}-\frac{d\ln\xi\eta^{4}}{dr}-\left(3+\xi\right)\frac{d\ln r_{c}}{dr}=-\frac{4r_{c}}{r^{2}}\label{eq:machgeneral}
\end{equation}
where, $\eta=r/r_{c}$, $\xi=v^{2}/C_{s}^{2}$, $r_{c}=GM_{\sun}/2C_{s}^{2}=R_{\sun}\frac{v_{\sun}^{2}}{4C_{s}^{2}}$
and in the derivation of Eq. (\ref{eq: soundgeneral}) the obvious
relation
\begin{equation}
\frac{d\ln r_{c}}{dr}=-\frac{d\ln C_{s}^{2}}{dr}.
\end{equation}
was used. It is straightforward to show that Eq.~(\ref{eq:machgeneral}) can be rewritten in the following form
\begin{equation}
\frac{d\ln\left(C_{s}^{2}{}^{\frac{5-3\alpha}{\alpha-1}}\xi\eta^{4}\right)}{dr}=0.\label{eq:sounnon5/3}
\end{equation}
Integration of the latter equation gives the general solution
\begin{equation}
C_{s}^{2}{}^{\frac{5-3\alpha}{\alpha-1}}\xi\eta^{4}=C_{*}^{2},\label{eq:solutsounnon5/3}
\end{equation}
here, $C_{*}^{2}$ is a positive integration constant. For the convenience of derivation, we define the following additional quantities
\begin{equation}
    \eta_{\alpha}\equiv C_{s}^{2}{}^{\frac{5-3\alpha}{4\left(\alpha-1\right)}}\eta=\frac{r}{r_{c\alpha}},
\end{equation}
where, 
\begin{equation}
    r_{c\alpha}\equiv\frac{r_{c}}{C_{s}^{2}{}^{\frac{5-3\alpha}{4\left(\alpha-1\right)}}}.
\end{equation}
Using these definitions, we derive the equation for $\eta_{\alpha}$ as
\begin{equation}\label{eq:eta_alphae}
    \frac{2}{\alpha-1}\eta_{\alpha}^{4}-\left(\frac{GM_{\sun}}{2r}\right)^{\frac{5-3\alpha}{\alpha+1}}\left(4+Dr\right)\eta_{\alpha}^{\frac{8}{\alpha+1}}+C_{*}^{2}=0,
\end{equation}
in combination with the algebraic relation
\begin{equation}
\xi\eta_{\alpha}^{4}=C_{*}^{2},\label{eq:solutsounnon5/3-2}
\end{equation}
and subject to the middle point condition that states that the sonic point $r_{*}$ exists within the radial domain above the starting surface (see below) $r>r_{\mathrm{SS}}\geq R_{\sun}$. The latter condition determines the position of the sonic point at $r=r_{*}$, $\xi_{*}=1$ and $\eta_{\alpha*}^{4}=C_{*}^{2}$, which leads to the general equation for $r_{*}$
\begin{equation}\label{eq:sonicpoint}
    \frac{\left (\alpha+1 \right)C_{*}^{2}{}^{\frac{\alpha-1}{\alpha+1}}}{\left (\alpha-1 \right)\left(\frac{GM_{\sun}}{2}\right)^{\frac{5-3\alpha}{\alpha+1}}}r_{*}^{\frac{5-3\alpha}{\alpha+1}}-Dr_{*}-4=0.
\end{equation}
The position of the sonic point is defined as a positive, real solution of Eq.~(\ref{eq:sonicpoint}) that falls within the radial domain defined above. It should be noticed that for any rational value of $\alpha$ Eq.~(\ref{eq:sonicpoint}) is easily reducible to a polynomial equation. Once the position of the sonic point is determined, Eq.~(\ref{eq:eta_alphae}) can be solved numerically at each sampling point of $r$, this way obtaining the transsonic radial profiles for the quantity $\zeta_{\alpha}=\eta_{\alpha}^{4}$ which obeys the equation
\begin{equation}\label{eq:zeta}
    \frac{2}{\alpha-1}\zeta_{\alpha}-\left(\frac{GM_{\sun}}{2r}\right)^{\frac{5-3\alpha}{\alpha+1}}\left(4+Dr\right)\zeta_{\alpha}^{\frac{2}{\alpha+1}}+C_{*}^{2}=0.
\end{equation}
From this solution of $\zeta_{\alpha}$ the radial profiles of all the physical quantities included in the model are then recovered. It is evident that the formalism presented here reduces to the case considered in \citet{Shergelashvili-etal-2020} by substituting $\alpha =5/3$. Therefore, this approach is a generalization to arbitrary polytropic indices \(\alpha>1\). Since for polytropic indices below 1, the model would correspond to such a strong heating that the  temperature would steadily increase with  heliocentric distance and this is not observed in stellar winds, that case is excluded.
\subsection{Sample solutions of the analytical model}
\begin{figure*}
	\centering
	\includegraphics[width=1\textwidth]{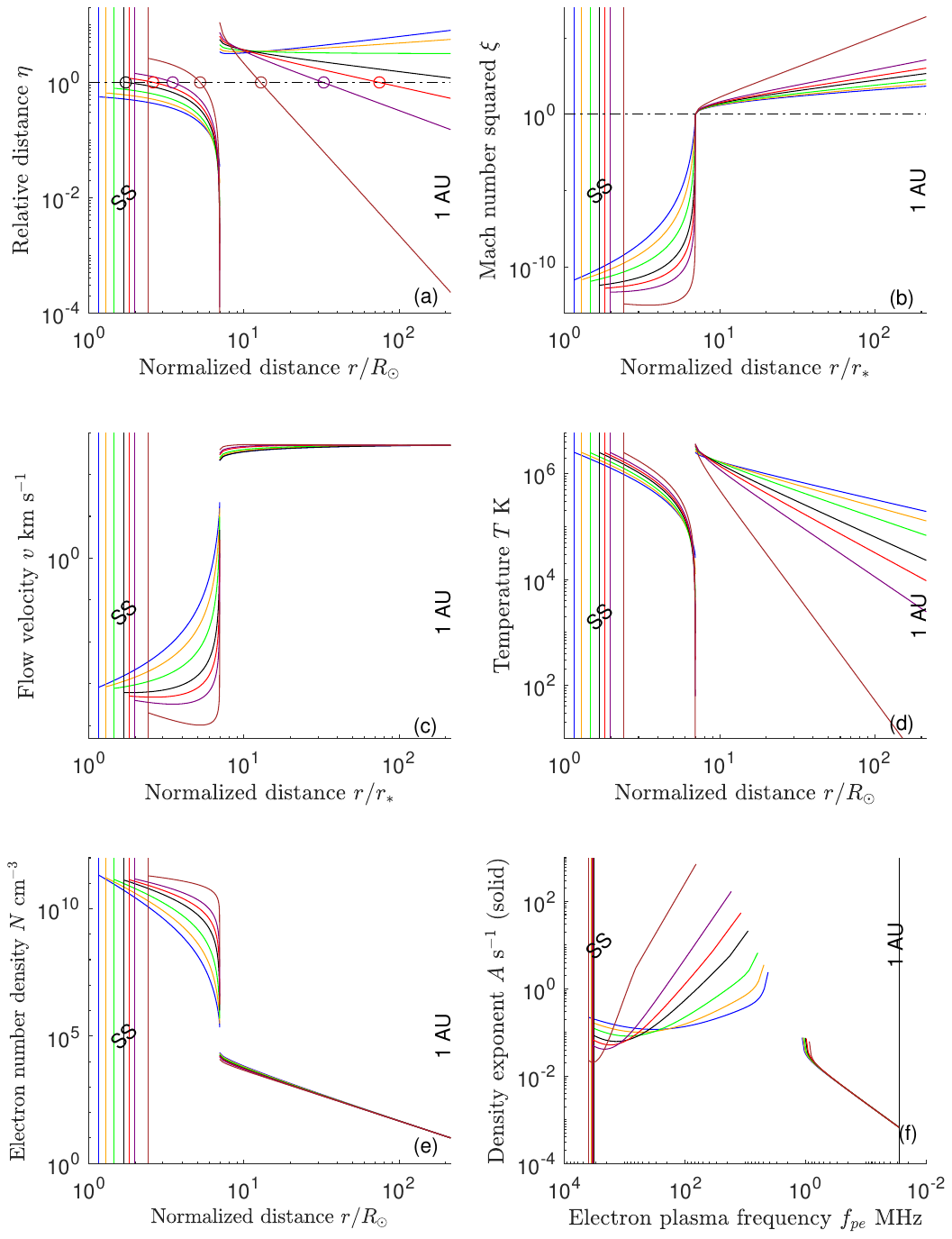}
    ~\vspace*{-2.0cm}\\
	\caption{Radial profiles of relative distance \(\eta\) (a), quadratic Mach number \(\xi\) (b), flow velocity \(v\) (c), temperature \(T\) (d) and number density \(N\) (e)  and the desinity exponent \(A\) against the electron plasma frequency \(f_{pe}\) (f) for a slow discontinuous solar wind stream and different polytropic exponents \(\alpha =4/3\text{ (blue)},\space 7/5\text{ (orange)},\space 3/2\text{ (green)},\space 5/3\text{ (black)},\space 9/5\text{ (red)},\space 2\text{ (purple)},\space 3\text{ (brown)}\).}\label{fig:SWanaslow}
\end{figure*}
\begin{figure*}
	\centering
	\includegraphics[width=1\textwidth]{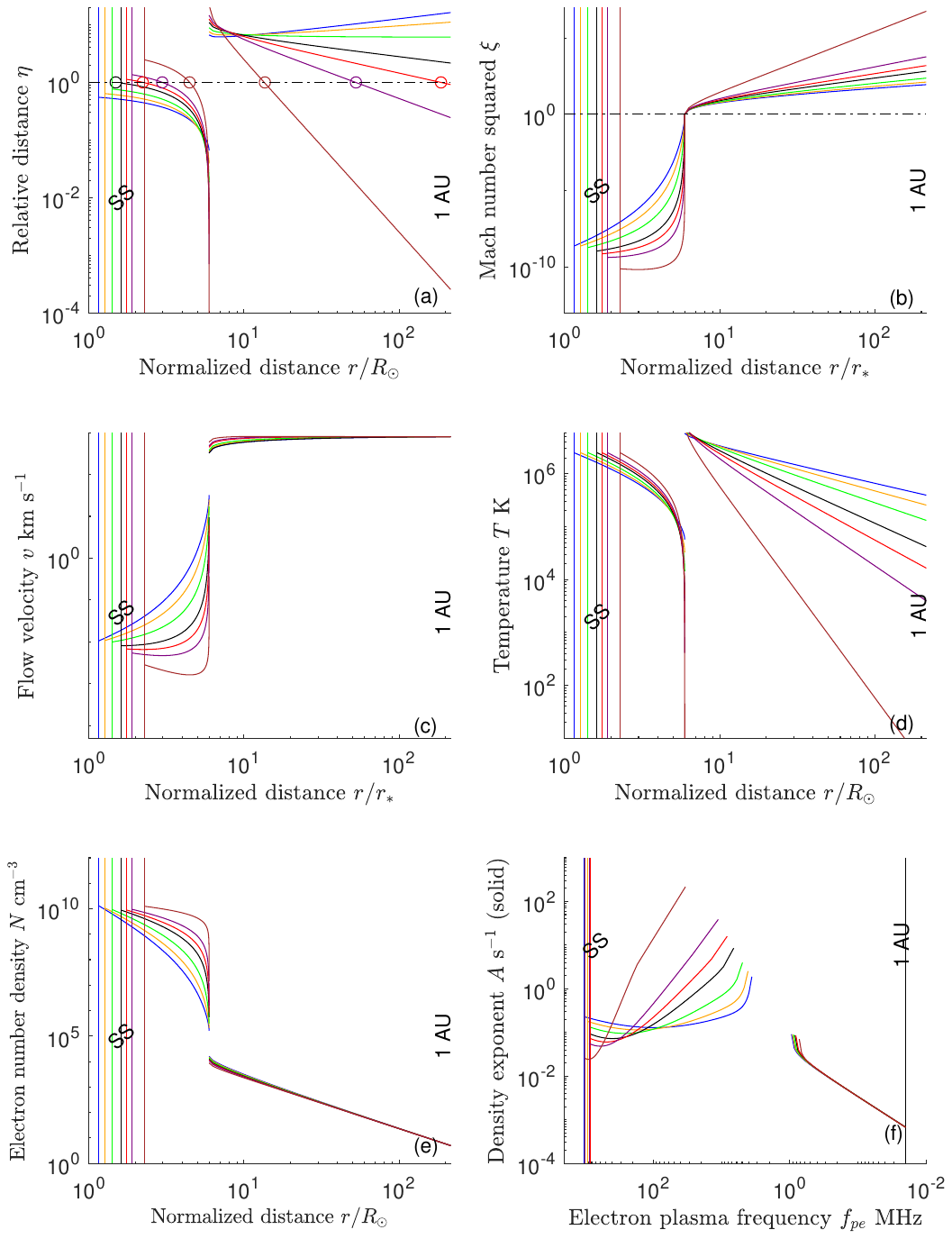}
    ~\vspace*{-2.0cm}\\
	\caption{Radial profiles of relative distance \(\eta\) (a), quadratic Mach number \(\xi\) (b), flow velocity \(v\) (c), temperature \(T\) (d) and number density \(N\) (e)  and the desinity exponent \(A\) against the electron plasma frequency \(f_{pe}\) (f) for a fast discontinuous solar wind stream and different polytropic exponents \(\alpha =4/3\text{ (blue)},\space 7/5\text{ (orange)},\space 3/2\text{ (green)},\space 5/3\text{ (black)},\space 9/5\text{ (red)},\space 2\text{ (purple)},\space 3\text{ (brown)}\).}\label{fig:SWanafast}
\end{figure*}
In figures \ref{fig:SWanaslow} and \ref{fig:SWanafast} we present the sample solutions for the cases of \(\alpha =4/3,\space 7/5,\space 3/2,\space 5/3,\space 9/5,\space 2,\space 3\) for a slow and a fast wind configuration (for the superadiabatic cases $\alpha > 5/3$ see the discussion in section~\ref{sec:heu}). For ease of comparison, we compute solutions with the same sonic point, which is $r_*=6.97R_{\sun}$ for the slow and $r_*=5.95R_{\sun}$ for the fast solar wind. This condition keeps all the input boundary conditions formerly used in case of $\alpha =5/3$ \citep[see tables in ][]{Shergelashvili-etal-2020} also $\alpha \neq 5/3$, except for the temperature values at the subsonic and supersonic boundaries. These have to be adjusted accordingly to ensure that the positions of the sonic point coincide in all the considered cases. 
It should be noted that this is an arbitrary choice for a better comparison between the different cases.\\
In addition, we only show the solutions down to a lower boundary where the temperature is \(2.5\cdot10^6K\). We refer to this lower boundary as the starting surface distance denoted with the subscript ($SS$).    Note that the solutions are, of course, mathematically constructable below the starting surface where they may, however, not be physically reasonable. Nonetheless, it is convenient for the following considerations to use both boundaries, i.e.\ the solar and the starting surface.\\
Upon examining the solutions displayed in Figs.~\ref{fig:SWanaslow} and \ref{fig:SWanafast}, one finds that the polytropic exponent \(\alpha\) has the strongest effect on the temperature. This is expected as it describes how temperature is related to the decrease in density \(\rho\) via
\begin{equation}
 \frac{\text{d}}{\text{d}r}\frac{T}{\rho^{\alpha-1}}=0.
\end{equation}
A lower polytropic exponent would lead to a less varying, higher temperature and a higher one to stronger cooling with heliocentric distance. Therefore, it appears that the jumps in the physical quantities are greater for higher polytropic exponents.\\
Polytropic exponents above \(5/3\) imply cooling of an expanding wind stream (see section \ref{sec:heu}) leading to partly decelerating winds. The crossings of the relative distance \(\eta\) of the unity line indicate the switch from acceleration to deceleration and vice versa. Note that the subsonic wind slightly decelerates for \(\eta>1\) and, in contrast, the supersonic wind for \(\eta<1\). However, an actually significant deceleration of the wind over an extended region is neither found from the model nor observed, so far, for the case of the Sun. \\
Furthermore, the flow speed in the subsonic region is very low. Therefore, there prevail 
nearly hydrostatic conditions. This and the decelerating supersonic wind can be understood from the fact that the total energy input is occurring at sonic point, which is in reality not the case, of course. There are many different momentum and energy sources below and above the sonic point, such as, e.g., damping of magnetohydrodynamic waves. Therefore, the solar wind plasma would be accelerated there.\\
Additionally, the analytically derived solutions are characterized by the jump discontinuities seen in Figures \ref{fig:SWanaslow} and \ref{fig:SWanafast}. However, it must be clearly stated that the discontinuity at the sonic point is a mathematical artifact of the stationary model, which does not include the explicit form of the local source of energy responsible for the appearance of these 'jumps'. As was first demonstrated in \citet{westrich2024} and as we will show in the section~\ref{sec:Numerical}, with an explicit and spatially somewhat extended source of local heating near the sonic point the discontinuity is replaced by sharp but continuous radial gradients. 
This is why we refer to the solutions as quasi-discontinuous. Another consequence of this mathematical artifact is that the model predicts precise values of physical quantities everywhere except for the sonic (jump) point. For all quantities undergoing an actual jump at the sonic point, the problem is numerically ill-posed at that point. As a result, the values of temperature, density, velocity, etc. at the jump point depend drastically on the numerical precision (i.e.\ on domain sampling rates). With the sampling rate we use, for example, the temperature value on the subsonic side goes as low as $10^1$~K (as shown in figures \ref{fig:SWanaslow} and \ref{fig:SWanafast}) at the sonic point, which is, of course, unrealistically low. While, thus, the delta function-like heating should not be considered as leading to particular predictions near the sonic point, it, nonetheless, suffices to illustrate the principal nature of the new solutions. 
Therefore, these analytical solutions could be used to calculate the radial profiles of the electron plasma frequency \citep{Shergelashvili-etal-2020}, which were found to be in good agreement with those  inferred from various observations including those of solar radio emission. Besides, we indicated in that work that an abrupt decrease of the decameter dynamic spectrum of so-called type III radio burst in the inner heliospheric region.  
The newly derived solutions in the present paper exhibit a stronger rarefaction for higher values of the polytropic exponent \(\alpha\) and a lower one for lower values leading to different abandoned frequency bands as shown in panels (f) of figures \ref{fig:SWanaslow} and \ref{fig:SWanafast}.
In this context, it is convenient to observe that the density depletions shown in panels (e) in figures \ref{fig:SWanaslow} and \ref{fig:SWanafast}, can be again \citep[like it was addressed in][]{Shergelashvili-etal-2020} interpreted as phenomenon linked with the presence of Type-III radio bursts with cut-off spectral drift profiles \citep{Melnik-etal-2014},  where the density exponent $A$ is defined as a constant independent of emission frequency $f_{pe}$, in the same way as in \citet{dididze2019} with characteristic local plasma probing electron beam velocities $v_c\approx 0.3c$, here $c$ is the speed of light. While in panels (f) it is clearly seen that assumption of $A$ being constant is a drastic idealization specifically for the upstream subsonic region. Instead, a plausible assumption has to be consideration of $A(f_{pe})$ as some nonlinear function of local radio emission frequency. This update can make possible the interpretation of other Type-III radio bursts with unusual spectral behavior, e.g. ones with spectral drift rate with changing sign \citep{Melnik2015}. Although, the latter issue deserves a special attention and has to be studied in a consistent manner. 

\section{Momentum and Energy Balance}\label{sec:EnergyBudget}
Before comparing the analytically derived solutions with implicit heating with new numerically derived solutions with explicit heating as done in \citet{westrich2024}, we discuss the energy budget that is needed to sustain a jump of the given magnitude and the question of momentum conservation at this jump. First, we provide an estimate of the energy flux density input required for the jumps. This was done for the case \(\alpha=5/3\) in \citet{Shergelashvili-etal-2020}, and the total energy flux density input required should be approximately the same for the cases \(\alpha\neq5/3\), due to the nearly identical boundary conditions. Therefore, we explicitly calculate these values. In \citet{westrich2024}, the heating strength was approximated by an estimate of the temperature jump resulting from the analytic method of \citet{Shergelashvili-etal-2020}. In this paper, we improve this by exactly determining how much energy flux density input is needed for the jump. This is required to derive the exact heating functions for each analytically derived solar wind solution and to obtain in this way better approximated numerically derived solutions in section \ref{sec:Numerical} that mimic the jump. After the derivation of the required energy flux density input needed for the jumps, we show that the energy density input is also manifested in the momentum flux equation but that these jumps do not violate the momentum conservation.\\
\subsection{Energy Budget}
We start with the energy conservation law inherited from the initial governing equations which looks as 
\begin{equation}
\frac{d}{dr}\left[\rho vr^{2}\left(h+\frac{v^{2}}{2}\right)\right]=0,
\label{eq:energyfluxcon}
\end{equation}
where 
\begin{equation}
h=-\frac{GM_{\sun}}{r}+\frac{C_{s}^{2}}{\alpha-1}.
\end{equation}
We can write it like

\begin{equation}
  \frac{1}{r^2}\frac{\mathrm{d}}{\mathrm{d}r}
  \left[
    Q_\rho\left(
      \frac{v^2}{2}
      +\frac{C_s^2}{\alpha-1}
      -\frac{GM_{\sun}}{r}
    \right)
  \right]
  = Q_{H,E}(r)
  \label{eq:energyfluxconQ}
\end{equation}
while adding the heating function \(Q_{H,E}(r)\) to the equation and using the mass flux conservation law \(Q_\rho=\rho vr^2=const.\) To specify the heating function, we recall the discussion at the end of the previous section. A finite total energy density flux \(C_{Fe}\) is added at the critical point and no other source is present. Consequently, the heating function can be best described as
\begin{equation}
    \label{eq:heatfuncdelta}
    Q_{H,E}(r)=C_{Fe}\delta(r-r_*).
\end{equation}
So, the heating strength is now fully described by the value \(C_{Fe}\). We can now write the energy conservation as
\begin{equation}
    \label{eq:energyfluxdiff}
    \text{d}\left(\frac{v^2}{2}+\frac{C_s^2}{\alpha-1}-\frac{GM_{\sun}}{r}\right)=\frac{C_{Fe}}{Q_\rho}r^2\delta(r-r_*)\text{d}r.
\end{equation}
This equation is now straightforward to integrate. However, one needs to be aware of the discontinuity caused by the delta function. Therefore, we divide the whole range into a subsonic and a supersonic region. The subsonic region is bounded by the solar surface at \(r=R_{\sun}\) (denoted with the subscript \(R_{\sun}\)) and by a point infinitesimally below the critical point \(r=r_1\simeq r_*\) (denoted with the subscript \(1\)) and the supersonic region by a point infinitesimally above the critical point \(r=r_2\simeq r_*\) (denoted with the subscript \(2\)) and the outer boundary at Earth orbit \(r=R_{AU}=1\text{AU}\) (denoted with the subscript \(AU\)). We can now integrate over the distinct regions to calculate exactly the energy density flux needed \(C_{Fe}\). First, we integrate between these two regions, that is, between \(r_1\simeq r_*\) and \(r_2\simeq r_*\). We obtain with
\begin{equation}
    \label{eq:Fe}
    F_e=Q_\rho\left(\frac{v^2}{2}+\frac{C_{s}^2}{\alpha-1}-\frac{GM_{\sun}}{r}\right)=vEr^2
\end{equation}
being the total energy flux density with the energy density E, the equation
\begin{equation}
    \label{eq:energyintegration12}
    C_{Fe}=\frac{1}{r_*^2}\left(F_{e,2}-F_{e,1}\right).
\end{equation}
With \(v_1=C_{s,1}\),\(v_2=C_{s,2}\) and \(Q_\rho=\rho_1v_1r_*^2=\rho_2v_2r_*^2\), we derive the equation
\begin{equation}\label{eq:CfeBid}
    v_1E_1+C_{Fe}=v_2E_2
\end{equation}
where 
\begin{equation*}
    E_{1}=\rho_{1}\left[\frac{C_{s,1}^{2}\left(\alpha+1\right)}{2\left(\alpha-1\right)}-\frac{GM_{\sun}}{r_{*}}\right]
\end{equation*}
and 
\begin{equation*}
E_{2}=\rho_{2}\left[\frac{C_{s,2}^{2}\left(\alpha+1\right)}{2\left(\alpha-1\right)}-\frac{GM_{\sun}}{r_{*}}\right]
\end{equation*}
which was also given in \citet{Shergelashvili-etal-2020}. Equation (\ref{eq:CfeBid}) can be considered as a jump condition similar to the Rankine-Hugoniot relations for shocks. However, it should be clear that the discontinuity here is not a shock because (i) a source of energy is needed to sustain the jump and (ii) the flow is accelerating and not decelerating.\\
Before calculating these values directly, it is beneficial to first estimate the required total energy flux density input and to  compare it to other characteristic energy flux densities of the system. With equation  (\ref{eq:CfeBid}) and the temperature values obtained from figures \ref{fig:SWanaslow} and \ref{fig:SWanafast}, it is straightforward to see that
\begin{equation}\label{estimationT1}
\frac{GM_{\odot}\left(\alpha-1\right)}{Kr_{*}\left(\alpha+1\right)}\gg T_{1}.
\end{equation}
where \(K=\alpha k_B/\mu m_p\) with the Boltzmann constant $k_B$ and $\mu m_p$ the mean particle mass expressed in terms of the proton mass. 
Physically, this means that in the subsonic state 1 the thermal energy density is lower than the gravitational one. However, the presence of the local energy source enables an overcoming of this difference and, accordingly, in the numerator (supersonic state 2) the thermal energy density exceeds that of gravity and there are no leading terms (both terms are of comparable orders of magnitude)
\begin{equation}\label{estimationT2}
\frac{GM_{\odot}\left(\alpha-1\right)}{Kr_{*}\left(\alpha+1\right)}\lesssim T_{2}.
\end{equation}
Consequently, it is easy to see, using equation (\ref{eq:CfeBid}) and the previous two estimates, that  
\begin{align}
\left|\frac{C_{F_{e}}}{v_{1E_{1}}}\right|
&=\left|\frac{v_{2}E_{2}}{v_{1}E_{1}}-1\right|
=\left|\frac{T_{2}-\frac{GM_{\odot}\left(\alpha-1\right)}{Kr_{*}\left(\alpha+1\right)}}{T_{1}-\frac{GM_{\odot}\left(\alpha-1\right)}{Kr_{*}\left(\alpha+1\right)}}-1\right|\nonumber\\
&\approx\left|-\frac{T_{2}-\frac{GM_{\odot}\left(\alpha-1\right)}{Kr_{*}\left(\alpha+1\right)}}{\frac{GM_{\odot}\left(\alpha-1\right)}{Kr_{*}\left(\alpha+1\right)}}-1\right|=\frac{Kr_{*}\left(\alpha+1\right)T_{2}}{GM_{\odot}\left(\alpha-1\right)}
\sim\Theta\left(1\right).\label{eq:Cfeestimate}
\end{align}
The latter estimate again confirms those given in \citet{Shergelashvili-etal-2020} and \citet{westrich2024}, namely that the additional energy flux needed to maintain the jump obtained at the sonic point is only as low as in the order of the gravitational energy of the plasma in that region. These conditions are quite easily realizable.\\
There is a slight problem with this way of calculating the energy flux input, because the values left and right of the jump are not unique but depend on how close to the jump they are calculated. Furthermore, one needs to calculate the analytically derived solutions to get the value \(C_{Fe}\). Therefore, for an improved calculation, the values evaluated around the sonic point should be replaced with well-defined ones. To this end, we integrate equation(\ref{eq:energyfluxdiff}) over the subsonic region and separately over the supersonic region to obtain the following equations
\begin{align}
    F_{e,1}=&F_{e,R_{\sun}}\label{eq:energyconsub}\\
    F_{e,2}=&F_{e,AU}.\label{eq:energyconsup}
\end{align}
Note that here we use the boundary condition at the solar surface (instead of at the starting surface) for better comparison (same inner boundary condition for number density and stream velocity).
With these equations we can substitute the values in equation (\ref{eq:energyintegration12}) evaluated in the sonic region with the values at the boundary. The value \(C_{Fe}\) can now be expressed as
\begin{equation}
    \label{eq:CFefinal}
    C_{Fe}=\frac{1}{r_*^2}\left(F_{e,AU}-F_{e,R_{\sun}}\right).
\end{equation}
The calculated values for the different polytropic indices \(\alpha\) and different energy density flux values are listed in Table \ref{tab:energyfluxcomparisionslow} for the slow wind example and in Table \ref{tab:energyfluxcomparisionfast} for the fast wind example, each calculated two times, namely via equation (\ref{eq:CfeBid}) and via  equation (\ref{eq:CFefinal}), to compare the two methods. 
\begin{table*}
    \centering
    \caption{Energy flux density values for different polytropic exponents for the slow wind scenario. The energy density flux needed for the jumps \(C_{Fe}\) is calculated by equation (\ref{eq:CFefinal}) using the boundary condition values (subscript \(B\)) and by equation (\ref{eq:CfeBid}) using the heating sheath values (subscript \(S\)) }
    \label{tab:energyfluxcomparisionslow}
    \begin{tabular}{c c c c c c c c c c}
        \hline \hline \\
        \(\alpha\)  & \(F_{e,R_{\sun}}/R_{\sun}^2\) &\(F_{e,SS}/r_{SS}^2\) & \(F_{e,1}/r_{*}^2\) & \(F_{e,2}/r_{*}^2\) & \(F_{e,AU}/(1AU)^2\) & \(C_{Fe,B}\) & \(C_{Fe,S}\) & \(\left|C_{F_{e,S}}/v_{1}E_{1}\right|\) & \(\left|C_{F_{e,S}}/v_{SS}E_{SS}\right|\)\\
        {      } &  \(gs^{-3}\) & \(gs^{-3}\) & \(gs^{-3}\) & \(gs^{-3}\) & \(gs^{-3}\) & \(gs^{-3}\) & \(gs^{-3}\) &  & \\ \hline 
        \(4/3\) & \(-5932.97\) & \(-4444.92\) & \(-123.16\)& \(644.90\) & \(0.68\) & \(765.82\) & \(767.84\) & \(6.23\) & \(0.17\)\\ 
        \(7/5\) & \(-6143.65\) & \(-3723.66\) & \(-127.50\) & \(623.48\) & \(0.65\) & \(748.77\) & \(750.76\) & \(5.89\) & \(0.20\)\\ 
        \(3/2\) & \(-6259.58\) & \(-2963.75\) & \(-129.89\) & \(607.37\) &\(0.64\) & \(735.07\) & \(737.04\) & \(5.67\) & \(0.25\)\\ 
        \(5/3\) & \(-6305.21\) & \(-2256.27\) & \(-130.83\) & \(597.68\) & \(0.63\) & \(726.34\) & \(728.29\) & \(5.57\) & \(0.32\)\\ 
        \(9/5\) & \(-6312.44\) & \(-1925.44\) & \(-130.98\) & \(595.29\) & \(0.62\) & \(724.10\) & \(726.05\) & \(5.54\) & \(0.38\)\\
        \(2\) & \(-6314.89\) & \(-1628.07\) &  \(-131.03\) & \(594.21\) & \(0.62\) & \(723.07\) & \(725.01\) & \(5.53\) & \(0.45\)\\ 
        \(3\) & \(-6315.48\) & \(-1090.81\) & \(-131.05\) & \(593.88\) & \(0.62\) & \(722.76\) & \(724.71\) & \(5.53\) & \(0.66\)\\
    \end{tabular}
\end{table*}
\begin{table*}
    \centering
    \caption{Energy flux density values for different polytropic exponents for the fast wind scenario. The energy density flux needed for the jumps \(C_{Fe}\) is calculated by equation (\ref{eq:CFefinal}) using the boundary condition values (subscript \(B\)) and by equation (\ref{eq:CfeBid}) using the heating sheath values (subscript \(S\))}
    \label{tab:energyfluxcomparisionfast}
    \begin{tabular}{c c c c c c c c c c}
        \hline \hline \\
        \(\alpha\)  & \(F_{e,R_{\sun}}/R_{\sun}^2\) &\(F_{e,SS}/r_{SS}^2\) & \(F_{e,1}/r_{*}^2\) & \(F_{e,2}/r_{*}^2\) & \(F_{e,AU}/(1AU)^2\) & \(C_{Fe,B}\) & \(C_{Fe,S}\) & \(\left|C_{F_{e,S}}/v_{1}E_{1}\right|\) & \(\left|C_{F_{e,S}}/v_{SS}E_{SS}\right|\)\\
        {      } &  \(gs^{-3}\) & \(gs^{-3}\) & \(gs^{-3}\) & \(gs^{-3}\) & \(gs^{-3}\) & \(gs^{-3}\) & \(gs^{-3}\) &  & \\ \hline 
        \(4/3\) & \(-5232.35\) & \(-3989.05\) & \(-148.50\) & \(1784.25\) & \(1.37\) & \(1934.67\) & \(1932.72\) & \(13.02\) & \(0.49\)\\ 
        \(7/5\) & \(-5565.24\) & \(-3500.13\) &  \(-157.89\) & \(1735.16\) & \(1.33\) & \(1894.91\) & \(1893.03\) & \(11.99\) & \(0.54\)\\ 
        \(3/2\) & \(-5777.80\) & \(-2908.05\) & \(-163.88\) & \(1699.08\) & \(1.30\) & \(1864.78\) & \(1862.94\) & \(11.37\) & \(0.64\)\\ 
        \(5/3\) & \(-5881.62\) & \(-2275.90\) & \(-166.81\) & \(1678.21\) & \(1.29\) & \(1846.81\) & \(1845.00\) & \(11.06\) & \(0.81\)\\ 
        \(9/5\) & \(-5903.00\) & \(-1950.71\) & \(-167.41\) & \(1673.33\) & \(1.28\) & \(1842.53\) & \(1840.72\) & \(10.99\) & \(0.94\)\\
        \(2\) & \(-5912.06\) & \(-1673.09\) & \(-167.67\) & \(1840.69\) & \(1.28\) & \(1840.69\) & \(1838.88\) & \(10.97\) & \(1.10\)\\ 
        \(3\) & \(-5915.16\) & \(-1150.50\) & \(-167.76\) & \(1670.67\) &\(1.28\) & \(1840.20\) & \(1838.40\) & \(10.96\) & \(1.60\)\\
    \end{tabular}
\end{table*}
The comparison shows that the differences between these two methods are very small, and thus negligible. However, equation (\ref{eq:CFefinal}) is more accurate and can be derived solely by well-defined boundary conditions and the critical distance \(r_*\). Additionally, the value \(C_{Fe}\) can be determined without calculating the analytical solution. We only need the values of the physical quantities at the boundary and the position of the discontinuity. Apart from the differences between the two equations, there is also no big difference in the calculated energy density flux needed for the jump for different polytropic exponent values.
Furthermore, we proved with tables \ref{tab:energyfluxcomparisionslow} and \ref{tab:energyfluxcomparisionfast} our above claim that the total energy flux density input is approximately the same for all polytropic indices.\\
Another aspect of our results is the relation of the additional energy flux to the energy flux rates at the sonic point (upstream side), at the starting surface, and at 1~AU. The values of the energy flux densities are shown in Tables~\ref{tab:energyfluxcomparisionslow} and \ref{tab:energyfluxcomparisionfast}. The values in the ninth column of these tables confirm our general claim based on estimations (see equation (\ref{eq:Cfeestimate})) that the extra energy flux required to maintain the process is of the order of digit multiples of the total energy flux at the sonic point in the upstream, subsonic side.\\
Lastly, we emphasize that the energy flux rates at 1~AU predicted by our model are in very good agreement with the solar wind energy flux measured by ULYSSES and WIND instruments (see, \citet{ChatGaetan2012}).\\
\subsection{Constraints due to momentum conservation}\label{subsec:momentum}
Now after we have seen that the energy budget is realistic for the conditions of the solar wind, we look at the momentum density flux. The strong acceleration at the discontinuities raises the question whether the momentum is conserved across this jump. 
Thus, we have to explain how a energy flux input \(C_{Fe}\) can cause a momentum increase across the jump. Therefore, we look at the momentum and energy conservation laws, which can be written as
\begin{align}
    \frac{1}{r^2}\frac{\text{d}}{\text{d}r}\left(\rho v^2r^2\right)+\frac{\text{d}p}{\text{d}r}&=-\rho \frac{GM}{r^2} \\
    \frac{1}{vr^2}\frac{\text{d}}{\text{d}r}\left[vr^2\left(\frac{1}{2}\rho v^2+\frac{\alpha}{\alpha-1}p\right)\right]-\frac{Q_E(r)}{v}&=-\rho \frac{GM}{r^2}.
\end{align}
These equations have the same right-hand side and, thus, we can derive after treating the left-hand sides as equivalent with the ideal gas law \(p=\rho KT\)
\begin{equation}
    Q_E(r)=\frac{K}{\alpha-1}v\rho\frac{\text{d}T}{\text{d}r}-KvT\frac{\text{d}\rho}{\text{d}r}.
\end{equation}
With definition (\ref{eq:heatfuncdelta}) of the heating function and the assumption that the physical quantities are constant before and after the jump, one can integrate and obtain with constant mass flux \(F:=\rho_1 v_1=\rho_2 v_2\) the following estimation:
\begin{equation}
    C_{Fe}=\frac{K}{2}\left[\frac{2F}{\alpha-1}\left(T_2-T_1\right)-\left(v_1T_1+v_2T_2\right)\left(\rho_2-\rho_1\right)\right].
\end{equation}
Please note that, because the integrand is not well defined at the discontinuity, this is an approximate but, nonetheless, still valid result of the integration. With the condition that the wind speed is equal to the sound speed on both sides respectively, i.e., $v_1=C_{s,1}=(\alpha KT_1)^{1/2}$ and $v_2=C_{s,2}=(\alpha KT_2)^{1/2}$, it is possible to derive a quartic equation for the variable $\Theta:=(T_2/T_1)^{1/2}$:
\begin{equation}
    \Theta^4+C_\alpha\Theta^3-\left(C_\alpha+C_H\right)\Theta-1=0
\end{equation}
with the constants \(C_\alpha:=\frac{3-\alpha}{\alpha-1}\) and \(C_H:={2C_{Fe}}/{(F K T_1)}\). This quartic equation has only one real positive solution for all reasonable values. For the case of no energy flux input this solution is equal to \(1\) and thereby indicating that no jump is possible without an additional energy density flux input. For the case of an input \(C_{Fe}>0\) the solutions are all above \(1\). This means that the temperature and therefore also the wind speed has to increase and consequently the density has to decrease. For the case of an energy flux density loss \(C_{Fe}<0\) the real positive solution lies between \(0\) and \(1\). Thus, the equation gives physical reasonable results. Therefore, this equation describe the limitation of possible jumps due to momentum and energy density flux conservation. However, this proves that these jumps do not violate the momentum and the energy density flux conservation.
\section{Numerical Validation of the analytically derived solutions}\label{sec:Numerical}
To verify the analytical results for arbitrary values of the polytropic index \(\alpha\), we used a similar model as in \citet{westrich2024}. With an exactly defined heating strength, we can now derive exact heating functions for each analytically derived solar wind solution that mimic the abrupt jump.
\subsection{Quasi-discontinuous solar wind model}
To model quasi-discontinuous solar wind streams, we are using the approach developed in \citet{westrich2024}
\begin{align}
\frac{dv}{dr} =& \frac{v}{r} \frac{2\gamma K T - K Q_{H,T}(r) r - \frac{G M_{\odot}}{r}}{v^2 - \gamma K T}\label{eq:velocitydiff} \\
\frac{dT}{dr} =& - (\gamma-1) T \left(\frac{2}{r} + \frac{1}{v}\frac{dv}{dr}\right) + Q_{H,T}(r)\label{eq:temperaturediff}
\end{align}
with
\begin{equation}
    Q_{H,T}(r)=\frac{\alpha-1}{CQ_\rho}r^2Q_{H,E}(r)
\end{equation}
being a modified version of the heating function for this model framework.
The strength of the heating function (\ref{eq:heatfuncdelta}) is now well defined with equation (\ref{eq:CFefinal}). As in \citet{westrich2024}, we substitute the delta function by a Gaussian function to reflect that the heating of the plasma does take place at one point but over an extended region: 
\begin{equation}
    Q_{H,E}(r)=\frac{C_{Fe}}{\sqrt{2\pi\epsilon^2}}\exp{\left(-\frac{\left(r-r_*\right)^2}{2\epsilon^2}\right)}
\end{equation}
where \(\epsilon\) determines the width of the heating region. In \citet{westrich2024} it was shown that the smaller $\epsilon$ the better  the analytically derived solutions are approximated. Thus, we set the width to a rather small value of \(\epsilon=1/8R_{\sun}\). 
This novel derived heating function allows us to model the point-like injection of energy at the critical point with even better agreement with the analytically derived solution than in \citet{westrich2024}.\\
To obtain steady-state solar wind solutions, we set a critical point heliocentric distance, calculated from that via equation (\ref{eq:velocitydiff}) the critical temperature and the critical velocity. Subsequently, as in \citet{westrich2024}, we integrated the differential equations (\ref{eq:velocitydiff}) and (\ref{eq:temperaturediff}) from the critical point in and outward to the boundaries of the system.\\
\subsection{Results}
\begin{figure*}
    \begin{center}
    \includegraphics[width=0.49\textwidth]{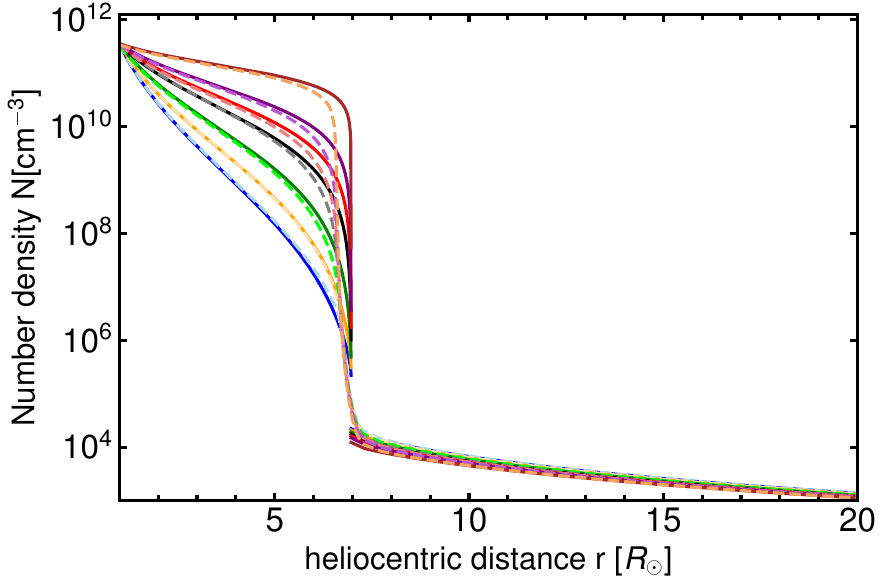}
	\includegraphics[width=0.49\textwidth]{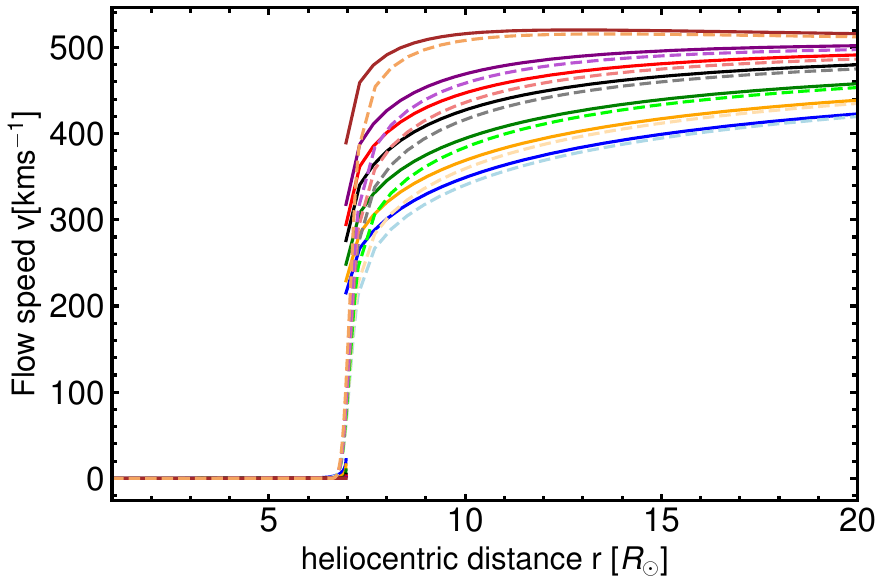}
	\includegraphics[width=0.49\textwidth]{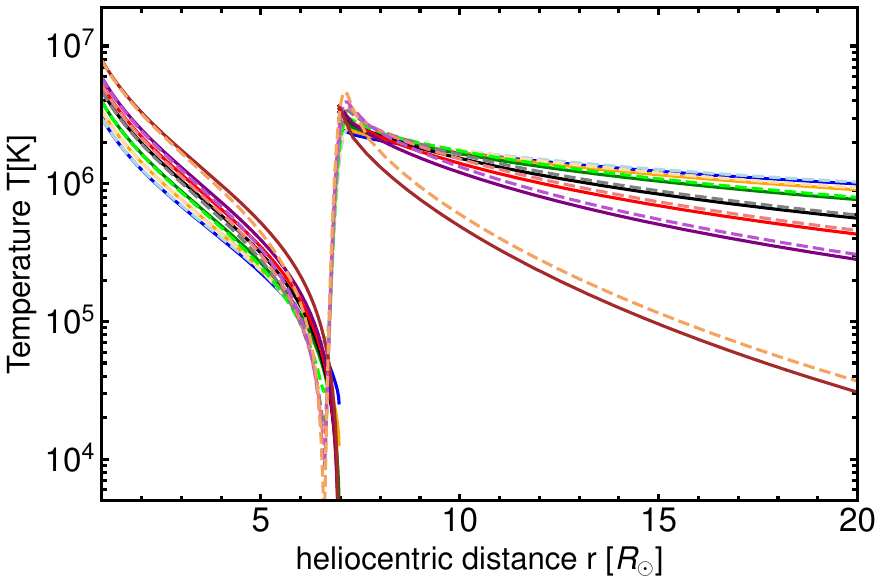}
	\includegraphics[width=0.49\textwidth]{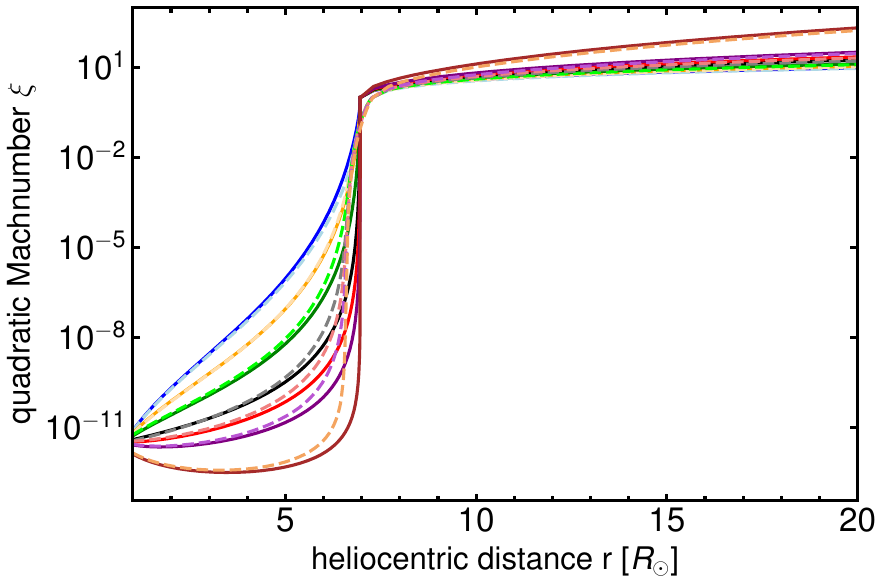}
    \end{center}
	\caption{Radial profiles (plotted between \(1\) and \(20R_{\sun}\)) of number density \(N\) (upper left), flow speed \(v\) (upper right), temperature \(T\) (bottom left) and quadratic Mach number \(\xi\) (bottom right) for a slow discontinuous solar wind stream (solid darker lines) and its quasi-discontinuous counterpart (dashed lighter lines) and different polytropic exponents \(\alpha =4/3\text{ (blue)},\space 7/5\text{ (orange)},\space 3/2\text{ (green)},\space 5/3\text{ (black)},\space 9/5\text{ (red)},\space 2\text{ (purple)},\space 3\text{ (brown)}\).}
    \label{fig:SWallslow}
\end{figure*}
\begin{figure*}
    \begin{center}
	\includegraphics[width=0.49\textwidth]{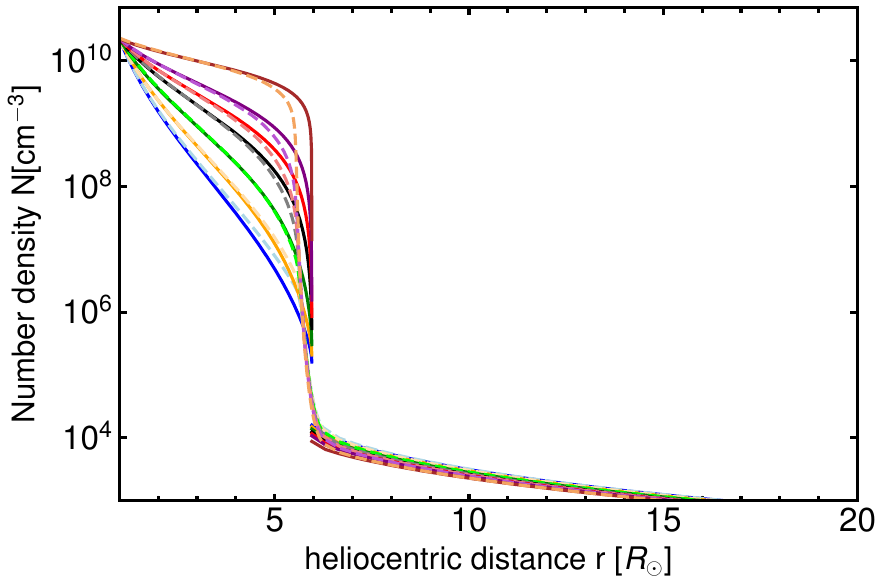}
	\includegraphics[width=0.49\textwidth]{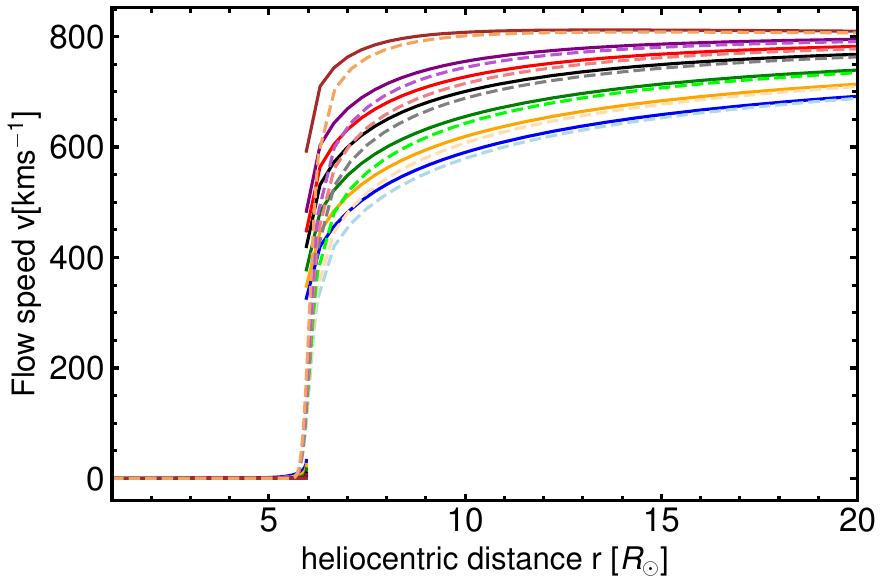}
	\includegraphics[width=0.49\textwidth]{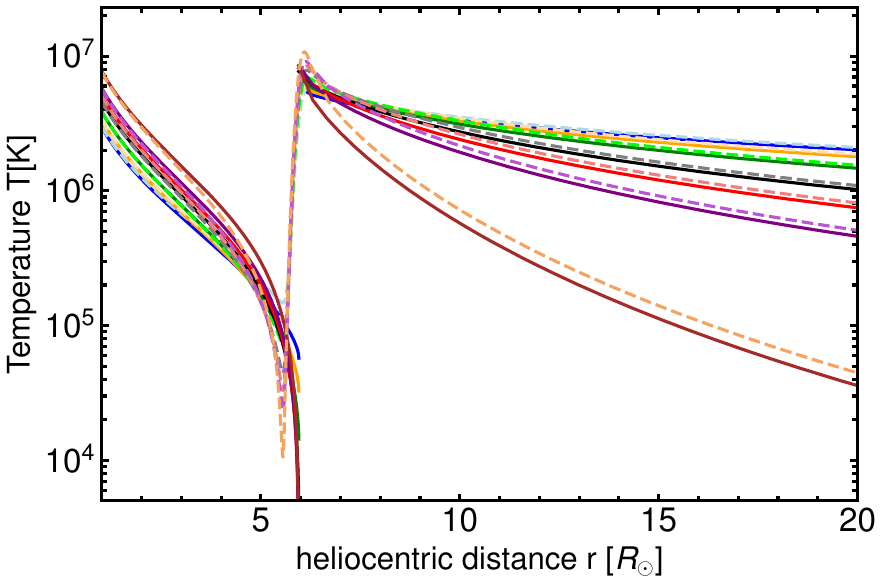}
	\includegraphics[width=0.49\textwidth]{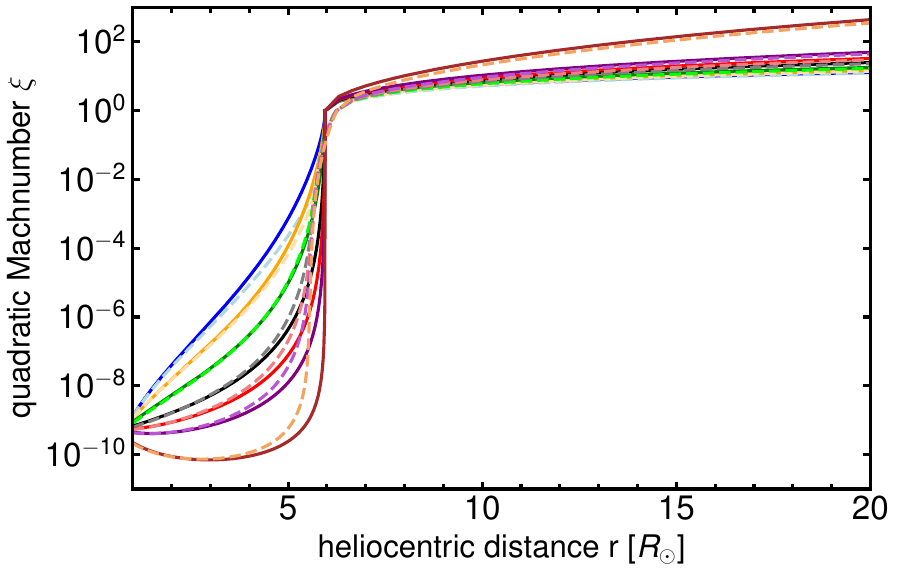}
    \end{center}
	\caption{Radial profiles (plotted between \(1\) and \(20R_{\sun}\)) of number density \(N\) (upper left), flow speed \(v\) (upper right), temperature \(T\) (bottom left) and quadratic Mach number \(\xi\) (bottom right) for a fast discontinuous solar wind stream (solid darker lines) and its quasi-discontinuous counterpart (dashed lighter lines) and different polytropic exponents \(\alpha =4/3\text{ (blue)},\space 7/5\text{ (orange)},\space 3/2\text{ (green)},\space 5/3\text{ (black)},\space 9/5\text{ (red)},\space 2\text{ (purple)},\space 3\text{ (brown)}\).}\label{fig:SWallfast}
\end{figure*}
Figures \ref{fig:SWallslow} and \ref{fig:SWallfast} display the approximation of the quasi-discontinuous solar wind model to the analytically derived solar wind solutions plotted on a linear distance scale between \(1R_{\sun}\) and \(20R_{\sun}\) for better comparison. As one can see, the differences  are very small. Only in the vicinity of the critical point are the differences more prominent. This is reasonable because we used a heating width of one-eighth solar radii, i.e.\ the heating region is small, but not a delta-function-like injection of energy. It is evident that for larger polytropic exponents \(\alpha\) the difference in this transsonic region is larger than for smaller ones. That can be explained by the larger jumps in the physical quantities for larger polytropic exponents. Nonetheless, the differences are small. Therefore, this numerical model proves that it is possible to construct with arbitrary polytropic indices \(\alpha>1\) a localized heating source that can produce a solar wind stream with steep gradients in physical quantities near the sonic point.\\
Furthermore, the improvements made in this paper, i.e.\ the explicit calculation of the required energy input, increased the agreement between the discontinuous and quasi-discontinuous solutions. That can be seen by looking at the stream velocity at the outer boundary. In \citet{westrich2024} the stream velocity in the supersonic region derived by the numerical model was a little lower than the analytically derived stream velocity in this whole region, because the required energy input was underestimated. The method greatly improved this, and now both models have the same solar wind velocities in the supersonic region, as can be seen in the velocity panels of figures \ref{fig:SWallslow} and \ref{fig:SWallfast}.\\
As discussed in \citet{westrich2024}, while the big temperature jumps at the sonic point in the continuous models pose the question of their 
stability against heat conduction or other temperature gradient-based processes, these processes do not drastically change the overall profiles, i.e.\ the characteristic high gradients in the vicinity of the sonic point persist. However, future studies should examine these effects more thoroughly.

\section{Heuristic grounds of the new solutions}\label{sec:heu}
After the above 
derivation of the model and the presentation of its solutions it appears helpful to provide a heuristic discussion of this modeling approach. In our first paper on this topic \citep{Shergelashvili-etal-2020} a basic reference model of discontinuous solar wind streams has been established. That model was developed within several simplifying assumptions, namely a steady-state radial outflow and an adiabatic expansion of a mono-atomic plasma outside the heating location. The latter two assumptions were also made in the follow-up work by \cite{westrich2024}, where the approach was extended to a finite, but still strongly localized, heating region and to an explicit time-dependence. With the additional assumption of $f=3$ degrees of freedom, and the usual relation $\gamma = 1 + 2/f$, the polytropic index is equal to the adiabatic one, i.e.\ $\alpha = \gamma = 5/3$. 

\subsection{The generalized polytropic index} \label{subsec:poly}

It is evident that particularly the last of the above assumptions represents a substantial idealization of the physical conditions in the solar wind. The inner heliosphere is densely populated with various stochastically distributed wave fields, and with dynamic processes maintaining states being in regimes of close or far from thermodynamic and/or statistical equilibrium. The presence of such entropy producing or dissipating processes as well as of macroscopic transport processes transforms the thermodynamic state of the system so that it does not depend solely on the microscopic degrees of freedom of the plasma particles, but also on other meso- and macroscopic thermodynamic processes. Such situation can be taken into account by introducing a generalized polytropic index that allows for the usual relationships  \citep{Chandrasekhar1933,Parker-1965,Livadiotis1_2018,Livadiotis2_2018}:
\begin{equation}\label{polytropic}
	p\propto N^{\alpha},T\propto N^{\alpha-1},
\end{equation}
but fulfills the following equation \citep{Livadiotis1_2018,Livadiotis1_2019,Livadiotis2_2019,Shaikh-etal-2023}:
\begin{equation}\label{work_heat}
	\frac{\delta w}{\delta q}=\frac{\gamma -1}{\gamma -\alpha},
\end{equation}
where the left-hand side is the ratio of pressure-volume work $\delta w$ done by the system on the surroundings and the total heat $\delta q$ transferred to the system. In this way the polytropic index is generalized to \citep{Livadiotis2_2019,Nicolaou-etal-2020,Shaikh-etal-2023}:
\begin{equation}\label{alpha_defin}
	\alpha =\frac{2}{f}\left ( 1-\frac{\delta q}{\delta w}\right )+1.
\end{equation}
This way, a subadiabatic $1<\alpha<\gamma=5/3$ or a superadiabatic $\alpha>\gamma=5/3$ behavior both in the young wind close to the Sun \citep[e.g., see][and references therein]{Nicolaou-etal-2020,Shaikh2025} and the 'Alf\'enic' wind at 1AU \citep[e.g., see][and references therein]{Shaikh-etal-2023,Shaikh2025} can be described. Note, that some in-situ observations at 1~AU even suggest $\alpha < 1$ 
\citep{Huang1989,Pudovkin2000}. The superadiabatic behaviour may become especially important in cases with strong plasma anisotropy (possibly resulting from wave heating), reducing the ability of charged particles to move perpendicular to the local magnetic field. Such reduction of the spatial degrees of freedom, not explicitly addressed in this paper, does not necessarily translate in a decrease of the effective degrees of freedom, or equivalently an increase of the polytropic index according to Eq. (\ref{alpha_defin}), as discussed, e.g., in \citet{Katsavrias2024L1}. It appears plausible that various parts of the solar wind encompass different physical conditions implying different (generalized) polytropic indices. So, a future generalization of the above model could allow for a spatially varying polytropic index. 

\subsection{Analogy with Laval nozzle engines}\label{subsec:nozzle}
The heuristics of the model have historically been based on the analogy of the solar wind governing equations with the model of Laval nozzle jet engines \citep{Clauser-1960, Shivamoggi-etal-2021}. The original Parker model was fitted directly to the engineering analog, demonstrating the fact that the pressure difference between the inner source surface (inlet nozzle) and the outer boundary at 1~AU (outlet nozzle) maintains the transsonic flow pattern (the sonic point exists in the system) when the critical gravitational escape point (throat of the nozzle) exists. In a first scenario, defined by the absence of  additional sources of energy along the path of the solar wind flow (inside the nozzle), the sonic point coincides with the gravitational escape point of the solar wind (equivalent to a coincidence of the sonic point with the nozzle throat), and the standard Parker's wind is formed. However, it is known that in the nozzles there are usually additional sources of fuel called afterburners, which enable the control of the extra power of the engine through the corresponding adjustment of the nozzle throat position \citep[being in accordance with the heuristic and physical grounds for an analogy with this type of afterburner injecting fuel/heat into the supersonic flow exhaust and providing extra thrust in Laval nozzle engines, introduced in the solar wind context by][]{Kopp-Holzer-1976}.
Analogously, the presence of the additional energy sources supplying the system in a stable way \citep[provided, e.g., by Alfv\`en wave turbulence][]{Hu-etal-1999,Tu-Marsch-1995,Vainio-etal-2003,Shergelashvili-Fichtner-2012}
and the resulting excess in energy (heat) flux along the path of the solar wind can lead to the formation of adjusted wind streams, which is a second scenario. Such adjustment may also be manifested by the occurrence of the temporary jumps at the sonic point if heat flux dissipates near that point \citep{Song-etal-2025}. 
Such jumps maybe smoothed by thermal conduction or other dissipation processes. 
Finally, there is a third scenario for the nozzle engines, when the flame ignited by the afterburner is able to have access to the fuel supplier so that more and more amount of fuels becomes being engaged into the burning process in a rather uncontrolled, explosive manner. Considering a similar situation in solar wind, when the source of additional energy (heat) flux, e.g., in form of an acoustic wave field born by some eruptive event, leads to jump-like steep gradients that engage more and more acoustic waves into the process so that the relaxation processes are overcome by the instability-governing 'acoustic afterburner' till the saturation level. In this case the relaxation back to the adjusted stable pattern of flow is impossible because the instability in the energy-supplying rate prevents the relaxation. This scenario is maintained as long as the unstable source of energy supply is in action. Once the excess 'fuel' is exhausted the relaxation process starts to prevail again and the jumps (or quasi-discontinuous profiles of the flow) disappear. Thus, such jumps may occur as temporary distortion of the solar (in general, stellar) wind. 


\section{Potential observational evidence for the new solutions}\label{subsec:swobs}
Besides the unusual type-III radio bursts \citep{Melnik-etal-2014}, whose cut-offs appear to be consistent with the strong density depletions as described above and in more detail in \citet{Shergelashvili-etal-2020}, one may speculate about other radiation signatures from that region. First, it should be noted that such 'radiation events' should occur only rarely. In case of the unusual type III radio bursts, a coincidence of various specific conditions is required. Obviously, the corresponding particle beam must cross the jump region to experience the effect of the strong plasma density depletion. However, the energy supply of the rapid acceleration event is both spatially and temporally localized and one must consider such events as being transient.

Second, and perhaps more fundamentally, it is unlikely that the drastic temperature variations in the vicinity of the sonic point are detectable in remote sensing measurements of EUV or white-light emission. To assess how much of the additional energy flux is diverted into radiative processes, we evaluate the thermal emission power on the subsonic and supersonic sides of the transition. As is well established, the solar corona - and particularly the sonic point region - is not sufficiently opaque to be characterized by black-body radiation. Instead, the emission follows a softer dependence on plasma temperature \citep{Rybicki1979, Aschwanden2005}: $P_\text{emm}\sim n^2 T^{1/2}$.
Using characteristic values for number density and temperature, we calculated from this expression the ratio of emission power immediately downstream the transonic jump (where the plasma temperature increases) relative to that upstream in the distance range from starting surface to region just before the jump and find ratios on the order which range in $10^{-15}-10^{-10}$ for the slow wind and $10^{-12}-10^{-8}$ for the fast wind. Furthermore, in the limit of closely approaching a sonic point, with the assumption made in subsection \ref{subsec:momentum} the ratio of the powers directly in the sonic point is \(P_{\text{emm},2}/P_{\text{emm},1}=n_2^2T_2^{1/2}/n_1^2T_1^{1/2}=\Theta^{-1}<1\) for all values of the additional energy density flux \(C_{Fe}>0\). Therefore, the emission after the jump is lower than before. This confirms that the additional energy flux is almost entirely converted into the bulk acceleration of the plasma, rather than being lost to radiative sinks.

There may also be potential evidence for the quasi-discontinuous solutions from in-situ spacecraft measurements. And one can indeed notice that the radial slopes of the number density on the supersonic side measured by the PSP \citep{Kruparova2023} are in very good agreement with profiles obtained by us in both the current study as well as in the original investigation \citep{Shergelashvili-etal-2020}. Another example is, the identification of intervals in the PSP measurements of the solar wind near the sonic critical point \citep{Cheng-etal-2024}, during which the near-subsonic wind exhibits a low density, an extremely low speed, and a low proton temperature. Obviously, these characteristics correspond to those of the new solutions upstream of the jump region. 

\section{Conclusions}
\label{sec:conclusions}
In this paper, we have generalized the modeling of stellar wind streams with strongly localized heating to include non-adiabatic expansions. By extending previous analytical and numerical studies to arbitrary polytropic indices ($\alpha \neq 5/3$), we have derived a more comprehensive framework that may be useful for the modeling of stellar winds, in general, and for understanding the variability of the solar wind observed by missions like the Parker Solar Probe, in particular. The main results of our investigation can be summarized as follows:
\begin{itemize}
    \item \textbf{Generalization of the Polytropic Index:} We successfully derived analytical solutions for stationary stellar wind models with strongly localized heating for arbitrary polytropic indices $\alpha > 1$. The study revealed how the polytropic exponent significantly influences the radial temperature profiles and the magnitude of the 'jump' in physical quantities at the sonic point; specifically, higher values of $\alpha$ result in stronger cooling and larger discontinuities.

    \item \textbf{Physical Plausibility of regions of density depletion and variability of its exponent:} The resulting solutions exhibit strong density depletions that vary with the polytropic index. These profiles offer a physical explanation for the cut-offs observed in the dynamic spectra of unusual solar Type-III radio bursts, linking the rarefaction of the plasma to specific polytropic behaviours. The density depletions observed in our solutions reinforce the link to Type-III radio bursts with cut-off spectral drift profiles \citep{Melnik-etal-2014}. However, our analysis reveals that treating the density exponent $A$ as a constant \citep[as done in previous studies like][and references therein]{dididze2019} is a drastic idealization, particularly in the upstream subsonic region. Instead, interpreting $A(f_{pe})$ as a nonlinear function of the local emission frequency for indivitual Type-III bursts, offers a more plausible framework. This refinement opens the possibility of modeling unusual spectral behaviors, such as Type-III radio bursts with changing signs of spectral drift rates \citep{Melnik2015}, which warrants further consistent investigation.

    \item \textbf{Energy Budget and Flux:} We performed a precise calculation of the energy budget required to sustain the discontinuous transitions. We demonstrated that the additional energy flux density needed is relatively low - comparable to the gravitational energy of the plasma in the heating region - and that the predicted energy flux at 1~AU aligns well with in-situ solar wind measurements from the ULYSSES and WIND spacecrafts. Furthermore, we showed that momentum conservation constrains physical quantities across the jump unambiguously.

    \item \textbf{Numerical Validation:} We validated the analytical 'discontinuou' solutions using a numerical model that employs a Gaussian heating function to simulate a spatially extended source. This comparison confirmed that the analytical discontinuities are mathematical artifacts representing physically realistic, steep continuous gradients. Furthermore, the improved energy budget calculations resolved velocity discrepancies noted in previous studies \citep{westrich2024,Shergelashvili-etal-2020}, resulting in a match between the analytical and numerical transsonic flows. 

    \item \textbf{Heuristic Interpretation:} The heuristic analysis draws on the analogy between the solar wind flow and a Laval nozzle engine equipped with an afterburner. In this context, the localized heating (possibly driven by acoustic waves) acts as a fuel source that creates adjusted wind streams \citep[in accordance with concepts outlined in][]{Kopp-Holzer-1976} or, in extreme cases, instability-driven profiles corresponding to the derived quasi-discontinuous solutions.
\end{itemize}

In summary, this study provides a robust theoretical tool for modeling stellar and solar wind streams that deviate from standard adiabatic approximations. It lays the foundation for future examinations of the sources, the effects, and possible observational indications of strong localized heating in the vicinity of the sonic point. These non-adiabatic, quasi-discontinuous solutions appear promising for interpreting the complex variability and acoustic wave presence detected in the near-Sun environment. Future work should further examine the stability of these steep gradients against rapid localized damping of acoustic wave fields under the action of viscosity, thermal conduction, and other dissipation processes.


\section*{Acknowledgements}
The work of L.W and B.S. was supported by Shota Rustaveli Georgian National Science Foundation grant for Fundamental Research – FR-23-10719. We acknowledge partial financial support for B.S. from the Ruhr-Universität Bochum (RUB) with in the VIP programme and from the Deutscher Akademischer Austauschdienst (DAAD) within EU fellowships for Georgian researchers, 2023 (57655523) – project Ref. ID. 91862684. We gratefully acknowledge also opportunity for B.S. to benefit from the visiting senior researcher position approved by \"Osterreichische Akademie der Wissenschaften at the  Graz Instit\"ute f\"ur Weltraumforschung (IWF) and financial support for L.W. from the Cusanuswerk via a scholarship programme. We also thank Viacheslav S. Titov for discussions regarding the momentum conservation of the jump solutions. Furthermore, we thank the anonymous reviewers for their time and effort in reviewing this manuscript and for their comments, which have strengthened our analysis.

\section*{Data Availability}
No new data was generated or analysed during this investigation. The data of the produced figures can be requested from the authors.
 



\bibliographystyle{mnras}
\bibliography{stability_lit} 








\end{document}